%% file: unsupDoA.tex
\documentclass[conference, 10pt]{IEEEtran}
\IEEEoverridecommandlockouts
\usepackage{cite}
\usepackage{amsmath,amssymb,amsfonts}
\usepackage{array}
\usepackage[caption=false,font=footnotesize]{subfig}
\usepackage{stfloats}
\usepackage{url}
\usepackage{graphicx}
\usepackage{textcomp}
\usepackage{xcolor}
\usepackage{tikz}
\usepackage{ellipsis}
\usetikzlibrary{calc}
\usetikzlibrary{decorations.pathreplacing,decorations.markings,shapes.geometric}
\usetikzlibrary{calc,patterns,angles,quotes,shapes,arrows.meta}
\usetikzlibrary{chains}

\usepackage{mathtools}
\usepackage{bm}
\usepackage{etoolbox}
\usepackage{scalerel}

\tikzstyle{block} = [draw, rectangle, 
    minimum height=4em, minimum width=4em]
\tikzstyle{input} = [coordinate]
\tikzstyle{output} = [coordinate]
\tikzstyle{pinstyle} = [pin edge={to-,thin,black}]

\usetikzlibrary{positioning}

\tikzset{radiation/.style={{decorate,decoration={expanding waves,angle=90,segment length=5pt}}}}

\usetikzlibrary{spy}
\usepackage{pgfplots}
\usepackage{wrapfig}
\usetikzlibrary{arrows,shapes}
\usetikzlibrary{positioning,shapes.callouts}
\usepgfplotslibrary{groupplots,dateplot}
\usetikzlibrary{patterns,shapes.arrows}
\pgfplotsset{compat=newest}

\def\BibTeX{{\rm B\kern-.05em{\sc i\kern-.025em b}\kern-.08em
    T\kern-.1667em\lower.7ex\hbox{E}\kern-.125emX}}

\input{def}

\newcommand{\wout}[1]{\textcolor{black}{#1}}

\usepackage{fancyhdr}

\fancypagestyle{cfooter}{ %
	\fancyhf{} 
	\cfoot{	\small \copyright 2022 IEEE. Personal use of this material is permitted. Permission from IEEE must be obtained for all other uses, in any current or future media, including reprinting/republishing this material for advertising or promotional purposes, creating new collective works, for resale or redistribution to servers or lists, or reuse of any copyrighted component of this work in other works.
}
	
}

\begin{document}

\title{Unsupervised Parameter Estimation using Model-based Decoder
\thanks{This work was supported by the Federal Ministry of Education and Research of Germany in the programme of “Souverän. Digital. Vernetzt.”. Joint project 6G-life, project identification number: 16KISK002}
}
\newcommand{\jbig}{$\mathcal{J}$}
\author{\IEEEauthorblockN{Franz Weißer, Michael Baur, and Wolfgang Utschick\\}
\IEEEauthorblockA{\textit{TUM School of Computation, Information and Technology, Technical University of Munich, Germany} \\
\{franz.weisser, mi.baur, utschick\}@tum.de}
}

\maketitle

\thispagestyle{cfooter}

\begin{abstract}
	In this work, we consider the use of a model-based decoder in combination with an unsupervised learning strategy for direction-of-arrival (DoA) estimation. 
	Relying only on unlabeled training data we show in our analysis that we can outperform existing unsupervised machine learning methods and classical methods. 
	The proposed approach consists of introducing a model-based decoder in an autoencoder architecture which leads to a meaningful representation of the statistical model in the latent space of the autoencoder.
	Our numerical simulations show that the performance of the presented approach is not affected by correlated signals and performs well for both, uncorrelated and correlated, scenarios. 
	This is a result of the fact, that, in the proposed framework, the signal covariance matrix and the DOAs are estimated simultaneously.
\end{abstract}


\input{intro}
\input{system}
\input{method}

\input{simulations}

\input{conclusion}

\bibliographystyle{IEEEtran}
  
\bibliography{IEEEabrv,mybib}

\end{document}

%% file: def.tex
\usepackage{array}
\usepackage[caption=false,font=footnotesize]{subfig}
\usepackage{stfloats}
\usepackage{url}
\usepackage{graphicx}
\usepackage{textcomp}
\usepackage{xcolor}
\usepackage{tikz}
\usepackage{ellipsis}
\usetikzlibrary{calc}
\usetikzlibrary{decorations.pathreplacing,decorations.markings,shapes.geometric}
\usetikzlibrary{calc,patterns,angles,quotes,shapes,arrows.meta}
\usetikzlibrary{chains}

\usepackage[utf8]{inputenc}

\usepackage{mathtools}
\usepackage{bm}
\usepackage{etoolbox}
\usepackage{scalerel}

\newcommand{\uca}{\mathrm{UCA}}

\newcommand{\T}{\mathrm{T}}
\renewcommand{\H}{\mathrm{H}}

\newcommand{\argmax}[1]{\underset{#1}{\arg\max}\;}

\tikzstyle{block} = [draw, rectangle, 
minimum height=4em, minimum width=4em]
\tikzstyle{input} = [coordinate]
\tikzstyle{output} = [coordinate]
\tikzstyle{pinstyle} = [pin edge={to-,thin,black}]

\usetikzlibrary{positioning}

\tikzset{radiation/.style={{decorate,decoration={expanding waves,angle=90,segment length=5pt}}}}

\usetikzlibrary{spy}
\usepackage{pgfplots}
\usepackage{wrapfig}
\usetikzlibrary{arrows,shapes}
\usetikzlibrary{positioning,shapes.callouts}
\usepgfplotslibrary{groupplots,dateplot}
\usetikzlibrary{patterns,shapes.arrows}
\pgfplotsset{compat=newest}

\def\BibTeX{{\rm B\kern-.05em{\sc i\kern-.025em b}\kern-.08em
		T\kern-.1667em\lower.7ex\hbox{E}\kern-.125emX}}
		
\definecolor{blue(pigment)}{rgb}{0.6, 0.2, 0.6}

%% file: intro.tex
\section{Introduction}

Direction-of-arrival (DoA) estimation is the process of finding the directions from which several signals impinge on an antenna array.
This estimation task has been a relevant research topic in array signal processing for many decades and is still prominent today~\cite{Krim1996, Ahmad2014}.
Some of the early algorithms like multiple signal classification (MUSIC)~\cite{Schmidt1986}, or more recent ones like the sparse recovery $\ell_1$-SVD \cite{Malioutov2005} and the sparse iterative covariance-based estimator (SPICE)~\cite{Stoica2011}, belong to the most employed methods in this field.
Their application range is wide, for example, radar, sonar, wireless communications, etc. 

In the case of correlated signals, MUSIC has degrading performance since it does not take into account the correlations between different signals. 
For a uniform linear array (ULA), e.g., spatial smoothing~\cite{Shan1985} has been introduced to circumvent this issue.
In~\cite{Wax1994}, the spatial smoothing was extended to uniform circular arrays (UCAs).
However, this extension is only valid for certain array geometries where the distances between the antennas along the circle is small enough.

The last decade experienced a significant push in the direction of machine learning in a lot of fields of signal processing. 
Multiple new methods for DoA estimation were proposed that make use of deep neural networks.
The authors in~\cite{Bialer2019} train a regression neural network to obtain the model order and the DoAs.
In~\cite{Merkofer2022}, parts of the MUSIC algorithm are augmented using neural networks.
A completely new structure which estimates the angles successively is proposed in~\cite{Barthelme2021b}.
An overview of the different kinds of methods is given in~\cite{Yuan2021}.

A drawback, which all of the described machine learning based methods have in common, is that all of them are trained in a supervised fashion. 
That means a labeled dataset with annotated angles for each sample needs to be acquired.
In most applications, however, the existence of labeled data cannot be assumed or is only possible with high additional cost.
An unsupervised learning strategy, in contrast, can work directly with measurements without prior annotating, which is an advantageous property.
In~\cite{Yuan2021}, a first idea of an unsupervised learning approach is introduced, which is based on a sparse signal representation. 
In this manuscript, we propose a different technique based on the maximum likelihood (ML) principle in combination with an autoencoder (AE) structure. 
In particular, the decoder of the AE incorporates the statistical signal model and the encoder is implemented as a neural network, that delivers estimates of the model parameters similar to~\cite{Yuan2021}. But using the framework of ML, the proposed method outperforms~\cite{Yuan2021} significantly.


The contributions in this paper are an unsupervised method for DoA estimation that is neither limited to a specific array geometry nor does it require the assumption of uncorrelated signals.
Furthermore, numerical simulations show that our method is able to outperform related unsupervised methods for DoA estimation in the described application cases.

%% file: system.tex
\section{System Model}

Assume that an antenna array which is equipped with $M$ antennas receives $K$ different signals from the far field.
The corresponding DoAs are denoted by $\bm{\theta} = [\theta_1,...,\theta_K]^\T$.
After the narrowband system model, the signal vector at snapshot $t$ can be expressed as
\begin{align}
	\bm{y}(t) = \bm{A}(\bm{\theta})\,\bm{s}(t) + \bm{n}(t), \quad t = 1,...,N
	\label{eq:system} 
\end{align}
where $\bm{s}(t)\in \mathbb{C}^{K}$ and $\bm{n}(t)\in \mathbb{C}^{M}$ are the signals sent by the users and additive noise, respectively. We assume that $\bm{s}(t) \sim \mathcal{N}_{\mathbb{C}}(\bm{0}, \bm{C_s})$ and $\bm{n}(t) \sim \mathcal{N}_{\mathbb{C}}(\bm{0}, \sigma^2\mathbf{I})$.
The matrix $\bm{A}(\bm{\theta}) \in \mathbb{C}^{M \times K}$ is the array manifold given as
\begin{align}
	\bm{A}(\bm{\theta}) = [\bm{a}(\theta_1),\,\ldots,\,\bm{a}(\theta_K)],
\end{align}
where $\bm{a}(\theta_k)$ denotes the array response to the incoming angle~$\theta_k$.


%% file: method.tex
\section{Proposed Autoencoder Structure}

\subsection{Stochastic Maximum Likelihood}


When estimating a parameter vector $\bm{\theta}$ based on some given observation $\bm{y}$, the maximum likelihood (ML) estimator can be formulated as
\begin{align}
	\hat{\bm{\theta}} = \argmax{\bm{\theta}} p(\bm{y} \mid \bm{\theta}).
\end{align}
In the field of DoA estimation, commonly two such ML estimators are distinguished depending on the assumptions made on~$\bm{s}$~\cite{Krim1996}.
If the estimator sees $\bm{s}$ as a deterministic but unknown parameter, we refer to this as deterministic ML (DML). 
On the other hand, if the estimator assumes the signals to be Gaussian with $\bm{s} \sim \mathcal{N}_{\mathbb{C}}(\bm{0}, \bm{C_s})$, the stochastic ML (SML) can be formulated.
In the latter case, the distribution $p(\bm{y}\mid\bm{\theta})$ is a zero-mean complex Gaussian distribution with covariance matrix
\begin{align}
	\bm{C_y} = \bm{A}(\bm{\theta}) \bm{C_s} \bm{A}(\bm{\theta})^{\text{H}} + \sigma^2\mathbf{I}, \label{eq:cov}
\end{align}
which follows directly from the system model in \eqref{eq:system}.
In general, the signal covariance matrix $\bm{C_s}$ is not known at the receiver. 
The same holds for the noise variance $\sigma^2$.
Only the $N$ snapshots $\bm{y}(t), t = 1,\ldots,N,$ are given, which are independent samples from the distribution $p(\bm{y}\mid\bm{\theta})$. 
Note that this only holds for fixed angles $\bm{\theta}$, which we assume to be constant during the $N$ snapshots.
Thus, the aim of SML is the maximization of the joint likelihood of the snapshots with respect to $\bm\theta$, $\bm{C_s}$, and $\sigma^2$, i.e.,
\begin{align}
	\max_{\bm{\theta},\bm{C_s},\sigma^2} \quad\prod_{t=1}^N \frac{1}{\pi^M\det(\bm{C_y})}\exp\left(-\bm{y}(t)^\H\bm{C_y}^{-1}\bm{y}(t)\right). 
	\label{eq:max_joint}
\end{align}
Instead of solving \eqref{eq:max_joint}, commonly the logarithm of the joint likelihood is maximized, which, after some reformulation steps, results in the minimization
\begin{align}
	\min_{\bm{\theta},\bm{C_s},\sigma^2} &\left[\text{ln}\left(\det \left(\bm{C_y}\right)\right)  + \text{tr}\left(\bm{C_y}^{-1} \hat{\bm{C}}_{\bm{y}}\right)\right],  \label{eq:ML2}
\end{align}
with the sample covariance matrix
\begin{align}
	\hat{\bm{C}}_{\bm{y}} = \frac{1}{N} \sum_{t=1}^{N} \bm{y}(t) \bm{y}(t)^\H.
\end{align}
Solving~\eqref{eq:ML2} is not straight forward since the likelihood function is not convex and, thus, can only be achieved with high complexity~\cite{Bialer2019}. 
Hence, it is not a practical option to consider.
Under the assumption of a good enough initial estimate, \cite{Barthelme2021a} utilizes a block gradient decent of \eqref{eq:ML2} to optimize the results of an initial DoA estimation stage.
But this can only serve as a refinement method and relies heavily on the initial estimate.

To provide a practical estimator of the SML solution we will introduce our method in the following section, which directly learns the solution to \eqref{eq:ML2} by optimizing the log-likelihood using an AE structure.

\subsection{Model-based Decoder}

An autoencoder (AE) is a machine learning structure, which uses an encoder that maps from an input space to a lower dimensional latent space.
A decoder maps the lower dimensional latent space back to the input space.
The idea is that these latent representations capture the most important information about the input samples in a compressed way.
In general, the latent variables cannot directly be interpreted.
In~\cite{Bengio2013}, an overview on representation learning is given, where the AE is suggested as a suitable candidate. Different approaches on interpreting the latent variables of communication scenarios are presented in~\cite{Weisser2021, Baur2022}.

\begin{figure}
	\centering
	\resizebox{0.5\textwidth}{!}{%
		\includegraphics[scale=1]{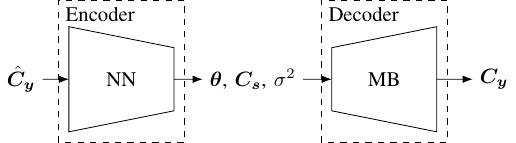}
	}%
	\caption{The proposed autoencoder structure with a model-based (MB) decoder according to \eqref{eq:cov}. A neural network as an encoder delivers estimates of $\bm{\theta}$, $\bm{C_s}$, and $\sigma^2$ based on which the covariance matrix $\bm{C_y}$ can be constructed.}
	\label{fig:arch2}
\end{figure}

The underlying idea of \wout{the} proposed approach is to force the latent variables of the AE to resemble the parameters of interest in~\eqref{eq:ML2}.
This can be done by exchanging the decoder part of the AE with a deterministic model-based function, which incorporates the statistical model of the data. 
Only the encoder neural network is part of the learning process. 
Fig.~\ref{fig:arch2} illustrates the proposed architecture.
It is possible to see that estimated values of $\bm{\theta}$, $\bm{C_s}$, and $\sigma^2$, are obtained from the encoder based on the sample covariance, which is used as input for the AE.
The construction of the covariance matrix $\bm{C_y}$ according to \eqref{eq:cov}, based on the estimated values, is used as the model-based decoder of the AE.
Using the likelihood from \eqref{eq:ML2} as a loss function, we can train a neural network to jointly estimate the angles, signal covariance matrix and noise variance without knowledge of their true values.
Consequently, the proposed method is unsupervised.
The benefit of using the model-based decoder architecture is the interpretability of the latent space.
Not only are the DoAs learned during this process. 
Also the signal covariance matrix is learned and can, therefore, be estimated after training of the AE.

For the learning of $\bm{C_s}$, we can use structural knowledge of the signal covariance matrix and reformulate it using the Cholesky decomposition as $\bm{C_s} = \bm{L} \bm{L}^\H$,
where  $\bm{L}$ is a lower triangular matrix. Using this decomposition, we ensure that the matrix $\bm{C}_{\bm{s}}$ is hermitian and positive-definite.
Thus, when considering correlated signals, the proposed method also learns the correlations of the signals by estimating the real and imaginary part of the lower triangluar matrix $\bm{L}$.
The output of the encoder can therefore be summarized as $\{\bm{\theta}, \bm{L}, \sigma^2\}$.

We will see in our simulations, that by learning and estimating the signal covariance matrix and the noise variance at the same time as the DoAs we find promising solutions to the SML problem in \eqref{eq:ML2} which lead to better DoA estimates than comparison methods.





\subsection{Covariance Matching}

In contrast to the proposed method, the unsupervised learning strategy in~\cite{Yuan2021} uses a sparse signal representation in combination with a similar loss as the $\ell_1$-SVD method~\cite{Malioutov2005}. Both methods do not maximize the likelihood but rather focus on an element-wise squared error optimization of the covariance matrix similar to
\begin{align}
	\min_{\bm{\theta},\bm{C_s}}  \left|\left|\hat{\bm{C}}_{\bm{y}} - \bm{A}(\bm{\theta}) \bm{C_s} \bm{A}(\bm{\theta})^{\text{H}}\right|\right|_\mathrm{F}^2. \label{eq:cov_match}
\end{align}
Our numerical simulations will demonstrate that ignoring the stochastic nature of the problem leads to inferior estimation results in contrast to the proposed method.

%% file: simulations.tex
\section{Numerical Simulations}

We want to focus in our simulations on the algorithm itself and not on the optimization of the employed antenna array.
Hence, we do not compare different array structures like in~\cite{Tan2014a, Qin2015, Pal2010, Han2014} but present the reader results for a uniform circular array (UCA).
Simulations for a uniform linear array led to similar results.
The choice of UCA is motivated by the fact, that methods for correlated signals tend to perform poorly for this array type.
For a UCA, we have the following array response
\begin{align}
	\bm{a}_{\uca}(\theta) = 
	\begin{bmatrix*}[c]
		\exp\left(-\mathrm{j}2\pi\frac{R}{\lambda}\cos(\theta)\right)\vspace{5pt} \\
		\exp\left(-\mathrm{j}2\pi\frac{R}{\lambda}\cos\left(\theta - \frac{2\pi}{M}\right)\right) \\
		\vdots \\
		\exp\left(-\mathrm{j}2\pi\frac{R}{\lambda}\cos\left(\theta - \frac{2\pi(M-1)}{M}\right)\right)
	\end{bmatrix*},
\end{align}
where $R$ and $\lambda$ denote the radius of the circular array and the electromagnetic wavelength, respectively.
For our simulations, we consider a ratio between radius and wavelength of~$1$. 
We also assume that the DoAs lie in the same plane as the UCA.

We evaluate all methods with data sampled from the system model in \eqref{eq:system}.
The DoAs $\theta_k$ are drawn from a uniform distribution between $0$ and $2\pi$ which spans the whole azimuthal field of view.
The powers of the signals are drawn from a uniform distribution between $-9$\,dB and $0$\,dB and are, subsequently, normalized to satisfy $\mathrm{tr}(\bm{C_s}) = 1$. 
Accordingly, we define the SNR as $\frac{\mathrm{tr}(\bm{C_s})}{\sigma^2} = \frac{1}{\sigma^2}$.
The number of antennas is $M=9$ and we use $N=100$ snapshots.

In the case where the incoming signals are correlated, i.e., $\bm{C_s}$ is not diagonal,
the resulting signal covariance matrix can be formulated as 
\begin{align}
	\bm{C_s} = \bm{\Lambda}^{1/2} \bm{C}_\rho \bm{\Lambda}^{1/2}, \label{eq:corr}
\end{align}
where $\bm{\Lambda}$ is a diagonal matrix with the powers of the individual signals on its diagonal. The matrix $\bm{C}_\rho$ denotes the correlation between the sources.
For $K=3$ signals, we define $\bm{C}_\rho$ as
\begin{align}
	\bm{C}_\rho =
	\begin{bmatrix}
		1 & \rho & \rho^2 \\
		\rho & 1 & \rho \\
		\rho^2 & \rho & 1 
	\end{bmatrix},
\end{align}
with $\rho \in [0,1]$ denoting the correlation coefficient. 

We compare \wout{the} proposed method with the classical approaches MUSIC~\cite{Schmidt1986}, $\ell_1$-SVD \cite{Malioutov2005}, and SPICE~\cite{Stoica2011}. 
We also compare to the unsupervised learning method from~\cite{Yuan2021}.
For the case of correlated signals, the rootMUSIC extension to UCA~\cite{Tewfik1992} with and without spatial smoothing (SpS) as introduced in~\cite{Wax1994} is simulated.
Unfortunately, the extension of rootMUSIC and SpS to UCA shows a weak performance. 
In the original paper, only particular angles are studied. 
When simulated with randomly drawn angles in the whole field of view the performance degrades drastically. 
This is observed independently of the condition $M \gg 2\pi R/\lambda$, which makes the extension of rootMUSIC and SpS to UCA not useful for our scenario.
The methods of SPICE, MUSIC, and $\ell_1$-SVD as well as the method from \cite{Yuan2021} are all based on a grid, which we set to $1200$ points in the range of $0$ to $2\pi$.

The performance of the different estimators are compared using the root mean square periodic error (RMSPE)
\begin{align}
	\mathrm{RMSPE} = \sqrt{\frac{1}{K} \sum_{k=1}^{K} \mathbb{E}_\theta\left[\left|\mathrm{mod}_{[-\pi,\pi)} \left(\theta_k - \hat{\theta}_k\right)\right|^2\right]}. \label{eq:rmspe}
\end{align}
The expectation over all possible angles is empirically estimated by $10^4$ Monte Carlo simulation runs.

\subsection{Implementation Details}


The architecture chosen for the encoder neural network in Fig.~\ref{fig:arch2} consists of $4$ convolution layers with channels $64$, $128$, $256$ and $512$. 
The kernel size is $3\times3$ for all layers and the stride is set to $2$ for all layers, except for the first. 
The padding is set to $1$ for all layers, except for the last.
Each of the convolutions is followed by a ReLU activation function. 
The output of the convolution layers is fed into one linear layer with size $512$.
The output layer maps to the appropriate dimension for the latent variables $\bm{\theta}, \bm{C_s}$, and $\sigma^2$.

Since the angles lie in the interval between $0$ and $2\pi$, a sigmoid function scaled by $2\pi$ is applied to the $K$ outputs corresponding to the angles. 
We assume that the powers of the users satisfy $\mathrm{tr}(\bm{C_s}) = 1$, hence, when considering uncorrelated signals we can use the softmax function on the $K$ outputs of the neural network, which estimate the signal powers. 
In the case of correlated signals, the outputs directly estimate the real and imaginary part of the lower triangluar matrix $\bm{L}$.
The last output of the network represents the learned noise variance which we force to be greater than zero by feeding the last element of the encoder output through the exponential function.

For \wout{the} proposed model-based decoder (MBD), three different versions are trained.
With \textit{diag} and \textit{full} we denote the assumption on the signal covariance matrix belonging to uncorrelated or correlated sources, respectively. 
In addition, we use two different loss functions, \eqref{eq:ML2} and \eqref{eq:cov_match}, which are denoted as \textit{ML} and \textit{Cov}, respectively.
The training data type is denoted as \textit{Uncorr} for uncorrelated signals and \textit{Corr} for correlated signals. In the latter case, the correlation coefficient $\rho$ is sampled uniform between $0$ and $1$.
This comparison is included to highlight the dependency of the performance on the assumption of correlated or uncorrelated signals.
All hyperparameters of the method from~\cite{Yuan2021} are chosen by means of a random search since not all of them were specified in the original work.

\subsection{Training phase}

During training of the proposed method samples are drawn from the system model in~\eqref{eq:system}.
The SNR is sampled uniformly in the interval $[-10\,\text{dB}, 30\,\text{dB}]$. 
We update the neural network weights with a learning rate of $10^{-4}$ using the \textit{Adam} optimizer~\cite{Kingma2014}.
For the training, $16 \cdot 10^4$ batches with batch size $256$ are used.
Since we draw new samples continuously during the training, every gradient descent step is done on new unseen data samples.
Using this learning method, the training of the model becomes robust to overfitting.
After this offline training phase, which has no constraint on time and only needs to be performed once, the learned neural network weights are fixed and no additional data is needed in the application phase.

\subsection{Simulation Results -- Uncorrelated Signals}
\begin{figure}
	\centering
	\resizebox{0.5\textwidth}{!}{%
		\includegraphics[scale=1]{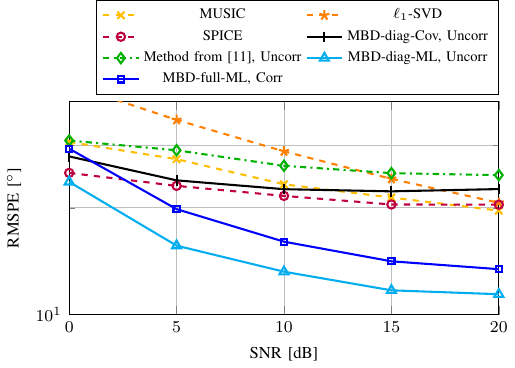}
	}%
	\\[-5mm]
	\caption{RMSPE over the SNR for $K=3$ uncorrelated signals.}
	\label{fig:init_ml}
\end{figure}

In Fig.~\ref{fig:init_ml}, the results for uncorrelated signals are presented. 
Both models which are trained with the SML loss function, dark blue and light blue line, outperform the classical methods and the method from~\cite{Yuan2021} for most SNR values. 
Additionally, it can be observed, that using \eqref{eq:cov_match} as the loss function (\textit{MBD-diag-Cov}) results in inferior performance, which could also be a reason why the method of~\cite{Yuan2021} does not perform as good as \wout{the proposed} technique.
Another observation made is that in the uncorrelated scenario the assumption on the matrix $\bm{C_s}$ being diagonal (\textit{MBD-diag-ML}) is beneficial for the estimation. 
This can be ascribed to the fact that the set of possible solutions is constrained using prior knowledge.
But even without using this prior information (\textit{MBD-full-ML}) the proposed method leads to better results than the comparison methods for SNR values greater then $5$\,dB.


\begin{figure}
	\centering
	\resizebox{0.5\textwidth}{!}{%
		\includegraphics[scale=1]{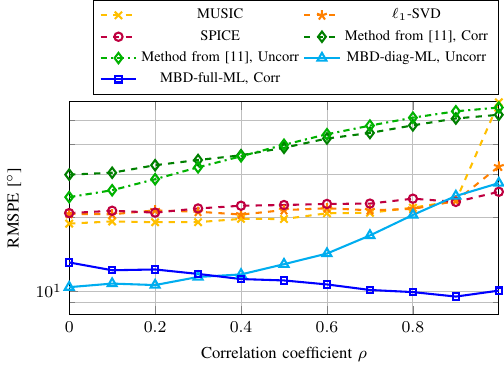}
	}%
	\\[-5mm]
	\caption{RMSPE for different correlation coefficients at an SNR of $20$\,dB for $K=3$ correlated signals.}
	\label{fig:init_corr}
\end{figure}

\begin{figure}
	\centering
	\resizebox{0.5\textwidth}{!}{%
		\includegraphics[scale=1]{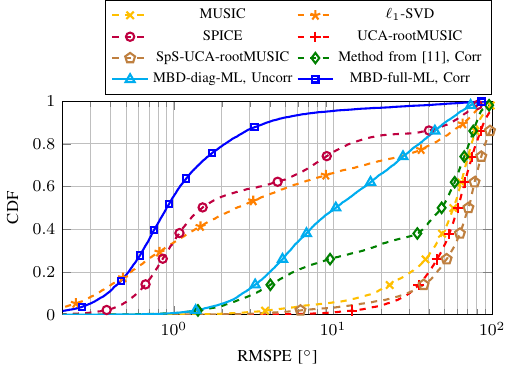}
	}%
	\\[-5mm]
	\caption{Empirical CDF of the RMSPE at an SNR of $20$\,dB for $K=3$ correlated signals with $\rho=1$.}
	\label{fig:init_cdf}
\end{figure}

\subsection{Simulation Results -- Correlated Signals}

Let us now relax the assumption of uncorrelated signals, which is a valid scenario when considering, e.g., multipaths.
In Fig.~\ref{fig:init_corr}, the RMSPE over the correlation coefficient is shown. 
It is visible that the comparison methods tend to degrade in their performances when the correlation coefficient increases.
Only SPICE achieves almost the same performance for fully correlated signals, i.e., $\rho=1$, as in the uncorrelated case.
\wout{The} proposed method using a full signal covariance matrix (\textit{MBD-full-ML}) slightly improves with increasing correlation. 
This can be explained by the fact that we estimate the correlation coefficient simultaneously with the DoAs.
In the case of full correlation, the performance gap in terms of the RMSPE is at its maximum.
This behavior of the estimation error for fully correlated signals is further highlighted in Fig.~\ref{fig:init_cdf} where the empirical cumulative distribution functions (CDFs) with respect to the RMSPE of all methods are illustrated.
In this figure, also the rootMUSIC algorithm for UCA and its spatial smoothening (SpS) variant are included.
For the proposed method, over $95\,\%$ of the test samples have an RMSPE lower than $10\,^\circ$, which means that very few outliers are present. 
In summary, we can conclude that the robustness to outliers is what makes \wout{the} proposed method with full signal covariance matrix perform so well for fully correlated signals.

%% file: conclusion.tex
\section{Conclusion}

In this work, we proposed a novel unsupervised DoA estimator that is superior to existing unsupervised machine learning techniques, as well as conventional estimators. 
The proposed model-based decoder does not only outperform the comparison methods for uncorrelated signals. 
It is also extremely powerful in the case of correlated signals.
Estimating all the signal/model parameters in the latent space of the autoencoder serves as a meaningful representation of the underlying statistical model. 
\wout{Since the method works completely independent of labeled training data, it is suitable for continuous improvement of a directional estimator in practical batch operation.}
Further investigations in our future work may include an extension to 3D angle estimation or removing the model order prerequisite, since in this work the model order is assumed to be given.